\documentclass[%
 reprint,
 amsmath,amssymb,
 aps,
]{revtex4-2}

\usepackage{graphicx}
\usepackage{dcolumn}
\usepackage{bm}

\begin{document}

\preprint{APS/123-QED}

\title{Balancing a vertical stick on a stochastically driven horizontal plate :\\ a variation on the Kapitza effect }

\author{Nachiketh M}
\email{nachikethm22@iitk.ac.in}
\author{J K Bhattacharjee}
\affiliation{%
Department of Physics, Indian Institute of Technology Kanpur, 208016 \\ 
 India
}%


\begin{abstract}
We consider the trick of balancing a vertical stick on a horizontal plate. It is shown that the horizontal stochastic driving of the point of contact can prevent the stick from falling provided that the stochasticity is that of a coloured noise with a correlation strength stronger than a critical value.
\end{abstract}

\maketitle
\section{\label{sec:level1}Introduction:\protect}
Stabilizing an unstable equilibrium point has been an interesting problem ever since Kapitza\cite{kapitza1965dynamic} showed how to stabilize an inverted pendulum by a very fast oscillation of the point of support. It is a non-intuitive result that has been explained carefully in classical mechanics texts\cite{LANDAU_1976} and pedagogical journals\cite{Blackburn1992StabilityAH}\cite{10.1119/1.1365403}. On a more formal basis, it has been explored by Levi\cite{doi:10.1137/1030140} and formal techniques of nonlinear dynamics in this context has been studied by Polekhin\cite{Polekhin2020TheMO}. Over the last decade, as techniques in control dynamics  have been extensively studied, there has been a lot of interest in different variants of the Kapitza pendulum\cite{Jones2016OpticalKP,Ahmad_2010,BHADRA2020110358,CIEZKOWSKI2021115761,7525341,10.1063/1.4890468,Polekhin_2023,Ramachandran_2023}. The widespread existence of the Kapitza phenomenon has been found even at the quantam scales over the last few years,\cite{Kuzmanovski_2024,Kulikov_2022,Golovinski_2022}, and extended to human walking very recently\cite{Reimann_2024}. Intermittent feedback control has also been developed as a control strategy for the inverted pendulum\cite{Morasso_2020} and was anticipated by Cabrera and Milton\cite{Cabrera_2002,Cabrera_2004} and Gawthorp etal\cite{Gawthrop_2013}. The focus, as expected, has been on the question of balancing a vertical stick - a popular fair-ground and classroom trick. In this work we show that for the vertical stick, a parametric modulation of the point of support on the horizontal surface can lead to stabilization if the modulation has the properties of a coloured noise. 

We begin by quickly recalling the Kapitza problem. In this set-up, a pendulum has its point of support vibrated vertically at a very high frequency \(\Omega\), and a small amplitude. This means that the natural frequency \(\omega\) of the pendulum will be subjected to oscillations of frequency \(\Omega\) and we can write the equations of motion as 
\[
{\ddot\theta} + \omega^2(1 + \epsilon \cos\Omega t) \sin\theta = 0   \label{1.1} \tag{1.1}
\]
For \(\epsilon=0\) the dynamics has two fixed points \(\theta^*\). One is \(\theta^* = 0\) which is stable and the other \(\theta^* = \pi\) which is unstable (upside down pendulum).
We consider small oscillations \(\delta\theta\) about the fixed points. Our interest is in the unstable fixed point, i.e \(\theta = \pi + \delta\theta\), which makes Eq\eqref{1.1}
\[
\delta{\ddot\theta} - \omega^2 (1 + \epsilon \cos \Omega t) \sin \delta\theta = 0   \label{1.2} \tag{1.2}
\]
Expanding, \(\delta\theta = \delta\theta_0 + \epsilon \delta\theta_1 + \epsilon^2 \delta\theta_2 + \ldots\), we obtain at O(1)
\[
\delta{\ddot\theta_0} - \omega^2 \delta\theta_0 = 0  \label{1.3a} \tag{1.3a}
\]
At O(\(\epsilon\)) and O(\(\epsilon^2\)) we obtain,
\begin{align*}
    \delta{\ddot\theta_1} - \omega^2 \delta\theta_1 - \omega^2 \delta\theta_0 \cos \Omega t = 0 \label{1.3b} \tag{1.3b} \\
    \delta{\ddot\theta_2} - \omega^2 \delta\theta_2 - \omega^2 \delta\theta_1 \cos \Omega t = 0 \label{1.3c} \tag{1.3c}  
\end{align*}
We are ignoring terms of O(\((\delta\theta_1)^2\))  in Eq\eqref{1.3c}. Since \(\omega << \Omega, \delta\theta_0\) is a slowly varying function and we can drop \(\delta\theta_1\) in comparison to \(\delta\ddot\theta_1\) to write 
\[
\delta\theta_1 \cong -\frac{\omega
^2}{\Omega^2} \delta\theta_0 \cos \Omega t \label{1.4} \tag{1.4}
\]
Inserting this \(\delta\theta_1\) in \eqref{1.3c} and keeping only the fast varying term, we have
\[
\delta\ddot\theta_2 \cong - \frac{\omega^4}{\Omega^2} \delta\theta_0 \cos^2 \Omega t \label{1.5} \tag{1.5}
\]

The equation of motion for \(\delta\theta\) is obtained as
\begin{multline*}
    \delta\ddot\theta = \delta{\ddot\theta_0} + \epsilon\delta{\ddot\theta_1} + \epsilon^2 \delta{\ddot\theta_2} + \ldots \\
    =\omega^2 (\delta{\theta_0} + \epsilon\delta{\theta_1} + \epsilon^2 \delta{\theta_2} + \ldots) + \epsilon\omega^2\cos \Omega t \delta\theta_0\\
    - \epsilon^2 \frac{\omega^4}{\Omega^2} \cos^2 \Omega t \delta\theta_0 + \ldots  \label{1.6} \tag{1.6}  
\end{multline*}
we do a long time averaging of \(\delta\theta\) to write
\[
\langle\delta\ddot\theta\rangle = \omega^2 \langle\delta\theta\rangle - \frac{\epsilon^2 \omega^4}{2\Omega^2} \langle\delta\theta\rangle \label{1.7} \tag{1.7}
\]
Note that \(\delta\theta_0\) has no oscillatory behaviour and hence the time average of the second term on the r.h.s of Eq\eqref{1.6} is zero. Further, for the same reason, the third term on the r.h.s of Eq\eqref{1.6} can be set to zero and correct to O(\(\epsilon^2\)), \(\epsilon^2\delta\theta_0\) can be replaced by \(\epsilon^2\delta\theta\).
It can easily be seen from the r.h.s of Eq\eqref{1.7}, that for \(\epsilon^2 \omega^2 > 2\Omega^2\), the second term dominates and hence the deviation \(\delta\theta\) oscillates around the unstable fixed point \(\theta_0 = \pi\).

\section{\label{sec:level2}The vertical Stick:\protect}
In this section, we focus on a vertical stick of mass \(m\) and length \(l\), being balanced on a flat surface against the force of gravity. We simplify it to a two dimensional problem, where the stick can only fall along the angle \(\theta\) as shown in Fig.1.
\begin{figure}
    \centering
    \includegraphics[width= 1.0\linewidth]{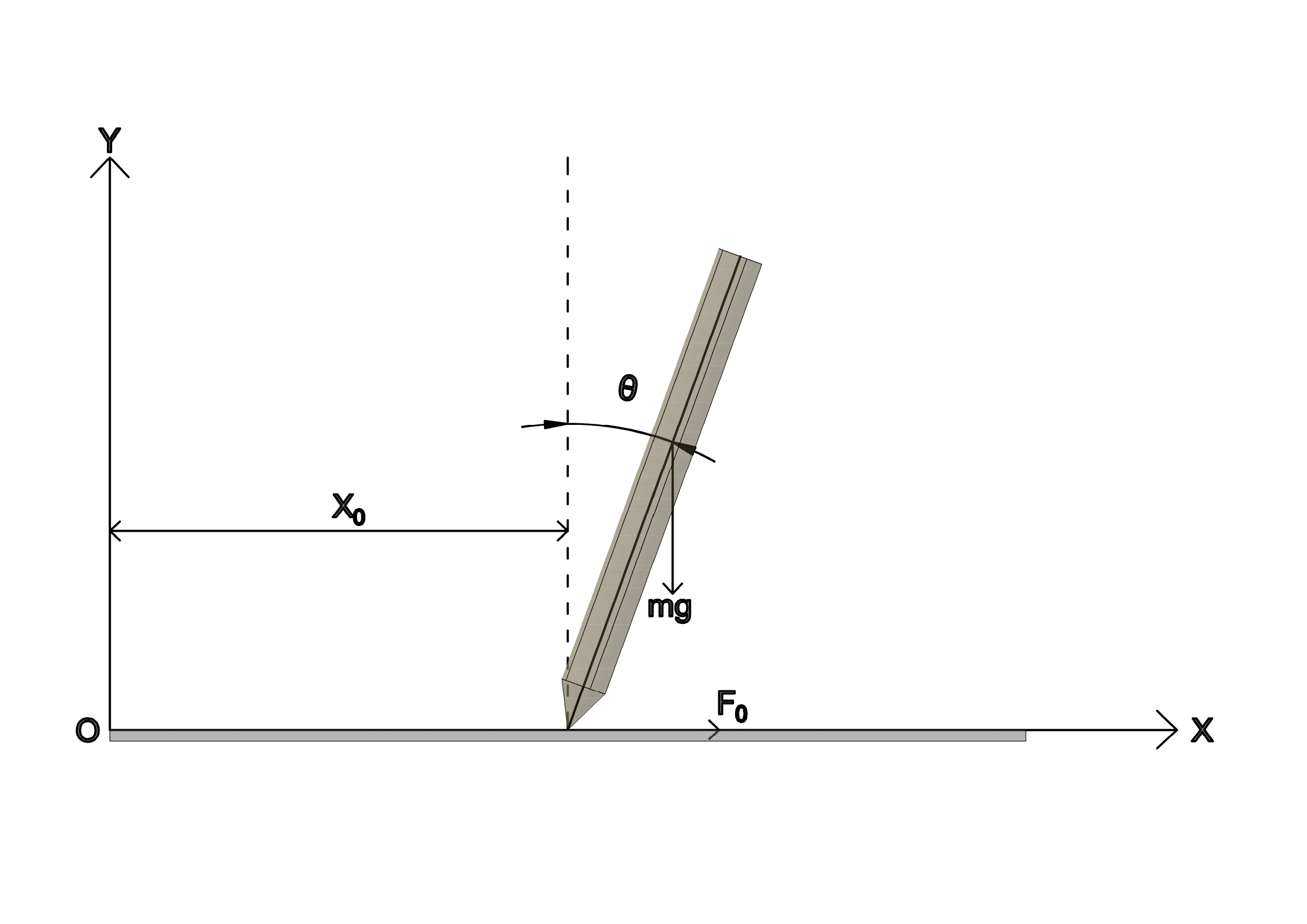}
    \caption{The vertical stick}
    \label{fig:enter-label}
\end{figure}

The initial \((t=0)\) position of the center of mass of the stick is \((x_0 + \frac{l}{2}\sin\theta, \frac{l}{2} \cos\theta)\) relative to the origin O of the co-ordinate axes as shown in Fig1. We apply a horizontal force \(F_0\) at the base of the stick to stabilize the falling stick for a small angular displacement \(\theta\). At any instant the y-coordinate of the center of mass will be \(\frac{l}{2}\cos\theta \) and the x-coordinate will acquire a time dependence as well, which makes the x-coordinate of the center of mass \(x(t) + \frac{l}{2}\sin\theta\). We introduce the horizontal and vertical coordinates \((X,Y)\) as  
\[
    X = x + \frac{l}{2}\sin\theta\\  \label{2.1a} \tag{2.1a}
\]
\[
    Y = \frac{l}{2}\cos\theta \label{2.1b} \tag{2.1b}
\]
The total kinetic energy is 
\begin{align*}
    K = \frac{m}{2} \dot x^2 + \frac{mL}{2}  \dot \theta \dot x \cos \theta + \frac{mL^2}{6} \dot\theta^2
    \label{2.2} \tag{2.2}
\end{align*}
The total potential energy is 
\[
V = \frac{mgl}{2} \cos\theta - F_0 x  \label{2.3} \tag{2.3}
\]
The Lagrangian of the system is:
\[
\frac{m}{2} \dot x^2 + \frac{mL}{2}  \dot \theta \dot x \cos \theta + \frac{mL^2}{6} \dot\theta^2 - \frac{mgl}{2} \cos\theta + F_0 x \label{2.4a} \tag{2.4a}
\]
The Equations of motion are obtained as
\[
\ddot x +   \frac{\ddot\theta l}{2}\cos\theta -   \frac{\dot\theta^2 l}{2}\sin\theta = \frac{F_0}{m}  \label{2.4b} \tag{2.4b}
\]
\[
\ddot x \cos\theta + \frac{2}{3} \ddot\theta l = g\sin\theta \label{2.4c} \tag{2.4c}
\]
Eliminating the variable x from Eq\eqref{2.4b} and \eqref{2.4c} takes us to
\begin{multline*}
    \ddot\theta[5-3\cos 2\theta] +  3 \dot\theta^2  \sin2\theta- 12\omega^2 \sin\theta = -12F\cos\theta   \label{2.4d} \tag{2.4d}
\end{multline*}

In the above equation \(F=\frac{F_0}{ml}\) and \(\omega^2 = \frac{g}{l}\)
The dynamics of Eq\eqref{2.4d} is understood by looking for its fixed points and its stability. It is seen that the only fixed point is \(\theta^* = \tan^{-1}(\frac{F}{\omega^2})\). To find the stability of the fixed point, we write \(\theta = \theta^* + \delta\theta\) and linearizing in \(\delta\theta\), we obtain
\[
\delta\ddot\theta = 6[\frac{\omega^2\cos\theta^* + F\sin\theta^*}{4 - 3\cos^2\theta^*}]\delta\theta  \label{2.5} \tag{2.5}
\]
The coefficient of \(\delta\theta\) is always positive making the fixed point unstable and the stick falls as expected. This raises the question whether the dynamics can be stabilized in the Kapitza fashion.

\section{\label{sec:level3}Kapitza like Stabilization: Perturbation Theory \protect}
We apply an Oscillatory/random force \(\epsilon F g(t)\) at the base of the stick in addition to the constant force \(F\) discussed in the previous section. Our aim will be to explore if a high frequency oscillatory \(g(t)\) or a random \(g(t)\) can help stabilize the stick. We begin rewriting Eq.\eqref{2.4d}, by including the time dependant force, as
\begin{multline*}
    \ddot\theta[5-3\cos 2\theta] + 3 \dot\theta^2 \sin 2\theta -12 \omega^2 \sin\theta \\ = -12F \cos\theta (1+\epsilon g(t))  \label{3.1} \tag{3.1} 
\end{multline*}
As shown in Sec.I, we expand the angle variable \(\theta\) as
\[
\theta = \theta_0 + \epsilon\theta_1 + \epsilon^2 \theta_2 +\ldots \label{3.2} \tag{3.2}
\]
At different orders of \(\epsilon\) the dynamics acquires the form
\begin{multline*}
    O(1):\qquad \ddot\theta_0[5-3\cos 2\theta_0] + 3\dot\theta_0^2 \sin 2\theta_0  - 12\omega^2\sin \theta_0 \\ = -12F\cos\theta_0  \label{3.3} \tag{3.3}
\end{multline*}
\begin{multline*}
    O(\epsilon):\qquad \ddot\theta_1[5-3\cos 2\theta_0] + 6\ddot\theta_0\theta_1  \sin 2\theta_0 + \\ 6 \dot\theta_0^2\theta_1  \cos 2\theta_0 + 6\dot\theta_0 \dot\theta_1 \sin 2\theta_0 - 12 \omega^2 \theta_1 \cos\theta_0  \\ = 12F[\theta_1 \sin\theta_0 - g(t) \cos \theta_0 ]
    \label{3.4} \tag{3.4}
\end{multline*}
\begin{multline*}
    O(\epsilon^2):\qquad \ddot\theta_2[5-3\cos 2\theta_0] + 6\ddot\theta_1 \theta_1 \sin 2\theta_0 \\ + 6\ddot\theta_0 [ \theta_1^2 \cos 2\theta_0 + \theta_2 \sin 2\theta_0] + 12 \theta_1 \dot\theta_1 \dot\theta_0 \cos 2\theta_0 \\ + 6\dot\theta_0^2[\theta_2 \cos 2\theta_0 - \theta_1^2 \sin 2\theta_0] + 3[\dot\theta_1^2 + 2\dot\theta_2 \dot\theta_0] \sin 2\theta_0 \\ + 6\omega^2[\theta_1^2 \sin\theta_0  - 2\theta_2 \cos \theta_0] \\ = 6F [\theta_1^2 \cos\theta_0 + 2 \theta_2 \sin\theta_0 + 2\theta_1 \sin \theta_0 g(t) ] \label{3.5} \tag{3.5}
\end{multline*}
At \(O(1)\), we have the dynamics studied in the previous section, and the outcome of that was the existence of an unstable fixed point \(\theta^* = \tan{^{-1}}(\frac{F}{\omega^2})\). This fixed point being unstable, we now proceed to higher orders, namely \(O(\epsilon)\) and \(O(\epsilon^2)\), to explore the effect of the modulating force \(g(t)\). At \(O(\epsilon)\), the dynamics is driven primarily by the rapidly changing \(g(t)\). This means that  terms with higher derivative dominate and the approximate solution of Eq.\eqref{3.4} can be obtained in the Kapitza fashion by considering the dominant part.
\[
[5-3\cos 2\theta_0] \ddot\theta_1 \cong -12F \cos\theta_0 g(t) \label{3.6} \tag{3.6}
\]
If \(g(t)\) is an oscillatory function \(A\cos\Omega t\), then 
\[
\theta_1(t) = \frac{12F \cos\theta_0}{5-3 \cos 2\theta_0} \frac{A}{\Omega^2} \cos \Omega t \label{3.7a} \tag{3.7a}
\]
On the other hand, if \(g(t)\) is a stochastic function, then the solution \(\theta_1(t)\) can be written as
\[
\theta_1(t) = - \frac{12F \cos\theta_0}{5-3\cos 2\theta_0} \int_{0}^{t} dt' \int_{0}^{t'} dt'' g(t'') \label{3.7b} \tag{3.7b}
\]
At the next order \(O(\epsilon^2)\), we need to identify the relevant terms that contribute to \(\ddot\theta(t)\) after an averaging over time has been carried out over the periodically/stochastically driven terms in Eq.\eqref{3.5}. This forces the relevant part of \(O(\epsilon^2)\) dynamics to be restricted to 
\begin{multline*}
    \ddot\theta_2[5-3\cos 2\theta_0] + 6\theta_1 \ddot\theta_1 \sin 2\theta_0 + 3 \dot\theta_1^2 \sin 2\theta_0 \\ = 12F \theta_1 \sin \theta_0 g(t)  \label{3.7c} \tag{3.7c}
\end{multline*}
We now need to construct the average values of the terms \(\theta_1\ddot\theta_1, \dot\theta_1^2\) and \(\theta_1 g(t)\) in order to arrive at the averaged equations of motion for \(\theta_2(t)\). To begin with we note from Eqs.\eqref{3.7a} and \eqref{3.7b}
that 
\[
\langle\theta_1(t)\rangle = 0 \label{3.8} \tag{3.8}
\]
Turning to Eq.\eqref{3.7c}, we note that on using Eqs\eqref{3.6},\eqref{3.7a} and \eqref{3.7b}, we can write the different contributions to \(\langle\ddot\theta_2\rangle\) as 
\[
\langle\theta_1\ddot\theta_1\rangle = -(\frac{12F \cos\theta_0}{5-3\cos 2\theta_0})^2 \frac{A^2}{2\Omega^2} \label{3.9a} \tag{3.9a}
\]
\[
\langle\theta_1\ddot\theta_1\rangle = (\frac{12F \cos\theta_0}{5-3\cos 2\theta_0})^2 (\frac{1}{\Gamma t}) \langle g(t)\int_{0}^{t} dt' \int_{0}^{t'} dt'' g(t'')\rangle  \label{3.9b} \tag{3.9b}
\]

Eq.\eqref{3.9a} above is for an oscillatory modulation and Eq.\eqref{3.9b} is for a random function \(g(t)\). Similarly, for an oscillatory function \(g(t)\) we have
\[
\langle\dot\theta_1^2\rangle = (\frac{12F \cos \theta_0}{5-3 \cos 2\theta_0})^2 \frac{A^2}{ 2\Omega^2}  \label{3.10a} \tag{3.10a}
\]
and for a random function \(g(t)\)
\[
\langle\dot\theta_1^2\rangle = (\frac{12 F \cos\theta_0}{5-3 \cos 2\theta_0})^2(\frac{1}{\Gamma t})\langle\int_{0}^{t} dt' g(t') \int_{0}^{t} dt'' g(t'')\rangle  \label{3.10b} \tag{3.10b}
\]

Finally, we have 
\[
\langle\theta_1 g(t)\rangle = \frac{6 F \cos\theta_0}{5-3 \cos 2\theta_0} \frac{A^2}{ \Omega^2} \label{3.11a}  \tag{3.11a}
\]
for the oscillatory \(g(t)\) and
\[
\langle\theta_1 g(t)\rangle = -(\frac{12 F\cos \theta_0}{5-3 \cos 2\theta_0})(\frac{1}{\Gamma t}) \langle g(t)\int_{0}^{t} dt' \int_{0}^{t'} dt'' g(t'')\rangle \label{3.11b} \tag{3.11b}
\]

It is apparent by substituting the averaged values that the oscillatory movement \(g(t) \propto \cos \Omega t\) cannot stabilize the stick. On the other hand, a stochastic \(g(t)\) has the potential for stabilizing the stick (the white noise situation where \(\langle g(t_1)g(t_2) \rangle \propto \delta(t_1 - t_2)\) does not give any contribution and does not stabilize the falling stick) if we use a coloured noise with
\[
\langle g(t_1)g(t_2) \rangle = A e^{-\Gamma|t_1 - t_2|} \label{3.12} \tag{3.12}
\]
Using the above definition the average values for the stochastic \(g(t)\) can be calculated as
\[
\langle\theta_1\ddot\theta_1\rangle = (\frac{12F \cos\theta}{5-3 \cos 2\theta})^2 (\frac{A}{\Gamma t})[\frac{1}{\Gamma^2} - \frac{1}{\Gamma^2} e^{-t\Gamma} - \frac{t}{\Gamma} e^{-t\Gamma}] 
\]

\[
\langle\dot\theta_1^2\rangle = (\frac{12F \cos\theta}{5-3 \cos 2\theta})^2 (\frac{A}{\Gamma^2})[2 + \frac{2}{\Gamma t}e^{-\Gamma t} - \frac{2}{\Gamma t} ] \label{3.13a} \tag{3.13a}
\]
\[
\langle\theta_1 g(t)\rangle = -(\frac{12F \cos\theta}{5-3 \cos 2\theta})(\frac{A}{\Gamma t}) [\frac{1}{\Gamma^2} - \frac{1}{\Gamma^2} e^{-t\Gamma} - \frac{t}{\Gamma} e^{-t\Gamma}] 
\]
in the large time limit, when \(t\rightarrow\infty\) we get
\[
\langle\theta_1\ddot\theta_1\rangle = 0
\]

\[
\langle\dot\theta_1^2\rangle = (\frac{12F \cos\theta}{5-3 \cos 2\theta})^2 \frac{2A}{\Gamma^2} \label{3.13b} \tag{3.13b}
\]
\[
\langle\theta_1 g(t)\rangle = 0
\]
We have
\begin{align*}
    \langle\theta\rangle = \langle\theta_0\rangle + \epsilon\langle\theta_1\rangle +
 \epsilon^2 \langle\theta_2\rangle + \ldots \\ 
 \langle\ddot\theta\rangle = \langle\ddot\theta_0\rangle + \epsilon\langle\ddot\theta_1\rangle +
 \epsilon^2 \langle\ddot\theta_2\rangle + \ldots \label{3.14} \tag{3.14}
\end{align*}
with,
\begin{align*}
    \langle\theta_1\rangle = 0 \\
    \langle\theta_2\rangle = 0  \\
    \langle\ddot\theta_1\rangle = 0           \label{3.15} \tag{3.15}
\end{align*}
So, after the averaging over the rapid oscillations is done we are left with
\begin{align*}
    \langle\theta\rangle & = \langle\theta_0\rangle \\
    \langle\ddot\theta\rangle & = \langle\ddot\theta_0\rangle + \epsilon^2\langle\ddot\theta_2\rangle  \label{3.16} \tag{3.16}
\end{align*}
Substituting Eq.\eqref{3.7a} to Eq.\eqref{3.13b} in \eqref{3.16} one gets
\begin{multline*}
    \langle\ddot\theta\rangle = - \frac{3\sin 2\theta}{5-3 \cos 2\theta} \langle\dot\theta^2 \rangle + \frac{12 \omega^2 \sin 
    \theta}{5-3 \cos 2\theta}\\ - \frac{12F \cos \theta}{5-3 \cos 2\theta} - \epsilon^2 [\frac{6 \sin 2\theta}{5-3 \cos 2\theta} \langle\ddot\theta_1\theta_1\rangle \\+ \frac{3 \sin 2\theta}{5-3 \cos 2\theta} \langle\dot\theta_1^2\rangle - \frac{12F \sin \theta}{5-3 \cos 2\theta} \langle\theta_1 g(t) \rangle]  \label{3.17a} \tag{3.17a}
\end{multline*}
\begin{multline*}
    \ddot\theta = \frac{12\omega^2 \sin\theta}{5-3\cos 2\theta} - \frac{12F \cos \theta}{5-3 \cos 2\theta} \\ -\epsilon^2[(\frac{3 \sin 2\theta}{5-3 \cos 2\theta})(\frac{12F \cos \theta}{5-3 \cos 2\theta})^2\frac{2A}{\Gamma^2} ]  \label{3.17b} \tag{3.17b}
\end{multline*}
Now expanding the above equation about the unstable equilibrium position for small \(\theta\) we get
\[
\ddot\theta = [6\omega^2 - \epsilon^2 A(\frac{108F^2}{\Gamma^2})]\theta - 6F \label{3.18} \tag{3.18}
\]
It is obvious that the above equation gives rise to oscillations when \(\epsilon^2 A( \frac{18F^2}{\Gamma^2} ) > \omega^2\). It is interesting to note that it is the combination \(\frac{F}{\Gamma}\) which determines the efficiency of the setup in effecting a stabilization. A small value of \(\Gamma\) corresponds to a strongly coloured noise i.e a noise which is correlated over a longer duration and that facilitates the stabilization.

\section{\label{sec:level4}Conclusion:\protect}
Balancing a vertical stick on one's hand against gravity has always seemed to provide a challenge to one's co-ordination of hand and eye. Over the last few years, this co-ordination has become the subject of generating control algorithms that can do the balancing act. In this work, we have cast this problem as an analogue of the Kapitza pendulum situation and shown that the oscillation will not work in this case. However , if the oscillation is replaced by a correlated coloured noise, then it is possible to obtain a stabilization over a critical strength of the noise-correlation. As expected, a longer correlation time for the noise leads to an easier stabilization.  

\bibliography{apssamp}

\end{document}